\documentclass[journal,article,accept,moreauthors,pdftex,10pt,a4paper,dvipsnames]{mdpimod}
\pdfoutput=1

\firstpage{1} 

\usepackage{float}
\usepackage{xargs}
\usepackage[]{xcolor} 
\usepackage[colorinlistoftodos,prependcaption,textsize=tiny]{todonotes}
\newcommandx{\unsure}[2][1=]{\todo[linecolor=red,backgroundcolor=red!25,bordercolor=red,#1]{#2}}
\newcommandx{\change}[2][1=]{\todo[linecolor=blue,backgroundcolor=blue!25,bordercolor=blue,#1]{#2}}
\newcommandx{\info}[2][1=]{\todo[linecolor=OliveGreen,backgroundcolor=OliveGreen!25,bordercolor=OliveGreen,#1]{#2}}
\newcommandx{\improvement}[2][1=]{\todo[linecolor=Plum,backgroundcolor=Plum!25,bordercolor=Plum,#1]{#2}}
\newcommandx{\reply}[2][1=]{\todo[linecolor=Gray,backgroundcolor=Gray!25,bordercolor=Gray,#1]{#2}}
\newcommandx{\av}[2][1=]{\todo[linecolor=Yellow,backgroundcolor=Yellow!25,bordercolor=Yellow,#1]{#2}}
\newcommandx{\thiswillnotshow}[2][1=]{\todo[disable,#1]{#2}}

\usepackage{subcaption}

\usepackage{tabularx}
\newcolumntype{b}{X}
\newcolumntype{s}{>{\hsize=\hsize}X}
\usepackage{array}

\newcommand{\cfr}{{\em cfr.}}
\newcommand{\ie}{{\em i.e.}}
\newcommand{\eg}{{\em e.g.}}

\newcommand{\R}{$R$}
\newcommand{\htlcsim}{{\bf CL}o{\bf TH}}

\newcommand{\payrate}{$r_{\pi}$}
\newcommand{\npay}{ $N_{\pi}$}
\newcommand{\npeers}{$N_p$}
\newcommand{\nchannels}{$N_{ch}$} 
\newcommand{\uncoopbefore}{$P_{\bar{c}_b}$}
\newcommand{\uncoopafter}{$P_{\bar{c}_a}$}
\newcommand{\gini}{$G$}
\newcommand{\sigmatop}{$\sigma_t$}
\newcommand{\sigmaam}{$\sigma_a$}
\newcommand{\capacity}{$C_{ch}$}
\newcommand{\samedest}{$F_{sr}$}

\newcommand{\psucc}{$P_s$}
\newcommand{\puncoop}{$P_{f_{\bar{c}}}$}
\newcommand{\pbalance}{$P_{f_b}$}
\newcommand{\proute}{$P_{f_r}$}
\newcommand{\punk}{$P_{\bar{k}}$}
\newcommand{\paytime}{$T$}
\newcommand{\attempts}{$N_a$}
\newcommand{\route}{$L_r$}
\newcommand{\ptimeout}{$P_{\bar{t}}$}

\newcommand{\simul}[1]{\texttt{\underline{#1}}}
\newcommand{\lnd}[1]{\texttt{#1}}
\newcommand{\msg}[1]{\texttt{#1}}
\newcommand{\node}[1]{$\mathcal{#1}$}

\newcommand{\payindex}{k}

\newcommand{\rt}{r_\payindex}
\newcommand{\bl}{b_\payindex}
\newcommand{\pay}{p_\payindex}
\newcommand{\graph}{\mathcal{G}}

\Title{\htlcsim: a Simulator for HTLC Payment Networks}

\titlenote{A previous version \cite{info9090223} of this work has been published on September 3 2018.}

\Author{Marco Conoscenti $^{1}$, Antonio Vetr\`{o} $^{1}$, Juan Carlos De Martin
  $^{1}$, \\
  Federico Spini$^{2}$, Fabio Castaldo$^{2}$ , Sebastiano Scr\`{o}fina$^{2}$}

\AuthorNames{Marco Conoscenti, Antonio Vetr\`{o}, Juan Carlos De Martin, Federico Spini, Fabio Castaldo, Sebastiano Scr\`{o}fina}

\address{%
  $^{1}$ \quad Nexa Center for Internet \& Society, Politecnico di Torino, DAUIN \\
  $^{2}$ \quad Fulgur Lab, correspondence \href{mailto:info@fulgurlab.com}{info@fulgurlab.com}
}

\lastrevision{\today}

\abstract{The Lightning Network (LN) is one of the most promising off-chain
  scaling solutions for Bitcoin, as it enables off-chain payments which are not
  subject to the well-known blockchain scalability limit. In this work, we
  introduce \htlcsim, a simulator for HTLC payment networks, of~which LN is the
  best working example. It simulates input-defined payments on an input-defined
  HTLC network and produces performance measures in terms of payment-related
  statistics, such~as time to complete payments and probability of payment
  failure. \htlcsim{} helps to predict issues that might arise in the development of an HTLC payment network, and to estimate the
  effects of an optimisation before deploying it. In upcoming works we'll
  publish the results of \htlcsim{} simulations.}


\copyrightyear{2018}

\begin{document}

\section{Introduction}

Bitcoin is a decentralised cryptocurrency that allows mistrusting peers
to send/receive monetary value without the need for intermediaries
\cite{nakamoto}. Bitcoin relies on the blockchain, a distributed
peer-to-peer public ledger where each peer stores all the history of
Bitcoin economic transactions. Its distributed and replicated nature
along with a limited block dimension entails capped transaction
\mbox{throughput \cite{accelerating}}. \mbox{As proven} in the
comprehensive study by Gervais et al. \cite{gervaisperformance}, a
commonly accepted blockchain design that resolves the scalability issue
is not yet known.

A way to address blockchain scalability is {\em off-chain scaling}
\cite{ln, decker, sprites, raiden, funding, khalil2017revive, flare},
which enables off-chain payments, \ie, payments that do not need to be
registered on the blockchain. {\em Payment channels} are key components
for off-chain scaling approaches. A payment channel is a two-party
ledger which is updated off-chain and uses the underlying blockchain
only as a settlement or dispute resolution layer. It allows an unbounded
number of off-chain payments to be sent/received between two involved
parties (the channel endpoints), as long as they can jointly reach
consensus.

Two-party payment channels can be linked together to build a {\em
  payment network} for off-chain payments. This allows parties not
directly connected by a payment channel to send/receive off-chain
payments which are routed across a network of linked payment channels.

The {\em Lightning Network} (LN) \cite{ln} is the most developed
off-chain solution for Bitcoin. The LN protocol leverages a contract
called {\em Hashed Timelock Contract} (HTLC) to build a payment network.
The HTLC scheme allows transferring off-chain payments through multiple
payment channels linked together. It guarantees transfer atomicity: even
if a payment is carried across multiple intermediate channels to reach
the recipient, the payment either succeeds or fails in all the involved
channels. Henceforth, we define an {\em HTLC payment network} as a
network of linked payment channels where payments are transferred using
the HTLC scheme.

At its current state of development, the LN protocol is characterised by
features that, if not properly understood, implemented and controlled,
might undermine the development of a healthy HTLC payment network, \ie,
a network that supports fast and successful payments. The features that
need to be studied are the following: (i) routing, \ie, a good routing
path that does not cause a payment failure depends on the knowledge of
an up-to-date network topology; (ii) channel capacity, which constrains
the payment amount; (iii) channel unbalancing, namely, the condition of
a skewed channel that has one endpoint depleted due to a number of
unidirectional payments (an issue not internally addressed by the LN
protocol); and (iv) uncooperative behaviour of peers involved in a
payment route, which increases payment time.

LN developers stress the importance of protocol improvements and
optimisation actions to guarantee correct functioning of the payment
network: in particular, they outlined the need for special routing
nodes, which are always online and contribute enough capital to route
payments \cite{lndev}. However, LN is an emergent network and the
resulting lack of central coordination does not allow to easily steer
network growth.

In this work we present \htlcsim, an HTLC payment network simulator. The
\htlcsim{} simulator could be used to: predict issues and obstacles that
might emerge in development stages of an HTLC payment network; assist in
planning uncoordinated development of the network; estimate the effects
of an optimisation before deploying it; predict the return on investment
of adding a hub to \mbox{the network}.

Examples of questions that can be answered by \htlcsim{} are: (i) ``How
many channels per peer are required to have a well connected network?'';
(ii) ``How do peers going offline affect performance?''; (iii) ``How
does payment amount influence the chance of payment failure?''; (iv)
``How does mean payment time decrease by adding a peer with a specific
set of payment channels in a specific section of the network?''.

The \htlcsim{} simulator is a discrete-event simulator that simulates
payments on HTLC payment networks. It takes as input parameters defining
an HTLC network (\eg, number of peers and number of channels per peer)
and parameters of payments to be simulated on the defined HTLC network
(\eg, payment rate and payment amounts). It generates performance
measures in the form of payment statistics, \eg, the probability of
payment failure and the mean payment complete time. To the best of our
knowledge, \htlcsim{} is the first available simulator for HTLC
\mbox{payment networks}.

The remainder of the paper is organised as follows.
Section~\ref{sec:background} is a background on Bitcoin, off-chain
scaling solutions and the Lightning Network. Section~\ref{sec:analysis}
describes the LN code used as reference to develop the simulator.
Section~\ref{sec:design} presents the \htlcsim{} simulator design.
Section \ref{sec:usability} reports about the usability of the
simulator. Finally, in Section~\ref{sec:conclusions} we draw the
conclusions of this work and outline possible future work.

\section{Background}
\label{sec:background}

In this Section, we briefly outline the background of our research:
Bitcoin and the blockchain; off-chain scaling solutions; and the
Lightning Network.

\subsection{Bitcoin}

Bitcoin is a decentralised payment system running on a peer-to-peer
network. Transactions are used to transfer bitcoin cryptocurrency and
they are defined by an input state and an output state. The output is
represented by the amount of currency to be transferred and by a Bitcoin
script defining the spending conditions to claim that amount. The input
is represented by the reference to an output of a previous transaction
and the proof fulfilling the spending conditions of the referenced
output. \mbox{The blockchain} is the distributed ledger technology
storing all Bitcoin transactions. Each peer (\mbox{that runs} a full
node) stores a whole copy of the ledger. Miners are special peers that
can write on the ledger: (i) they validate transactions broadcast on the
peer-to-peer network against the ledger in order to prevent double
spending of coins; and (ii) they gather transactions and write on the
ledger, by broadcasting a block (\ie, a collection of transactions),
statistically once every 10 min. For more details on Bitcoin and the
blockchain, we refer the reader to the detailed background study in
\cite{sok}.

Bitcoin's last year average throughput is slightly below 2.7 confirmed
transactions per second\footnote{Data fetched from
  \url{https://www.blockchain.com/en/charts/n-transactions} on November
24, 2018.}. In its current form, Bitcoin will hardly scale
beyond 100 transactions per second because of storage, processing,
latency, and bandwidth constraints \cite{decker}. For Bitcoin to become
a mainstream medium of exchange, transaction throughput should grow
significantly.

\subsection{Off-Chain Scaling Approaches}

Off-chain scaling addresses scalability by moving as many transactions
as possible off the blockchain.

\subsubsection{Payment Channels}

Payment channels \cite{paychannel} are the building block of an
off-chain payment network. Such channels are implemented as two-party
ledgers that allow two parties (the channel endpoints) to update their
balances off-chain, without the limit imposed to blockchain
transactions. The blockchain is only used as a dispute resolution layer
if and when consensus on the off-chain state cannot be reached between
channel endpoints.

In the rest of this section we'll use the following example of a \mbox{payment channel}.

\vspace{6pt}
\noindent {\em Payment channel example}. Alice and Berto open a payment
channel. Alice funds the channel with 0.5 BTC, Berto funds the channel
with 0.5 BTC, so Alice's initial {\em balance} in the channel is 0.5 BTC
and Berto's is 0.5 BTC. The {\em capacity} of the channel is the total
amount of bitcoins in the channels: \mbox{1 BTC}. After opening the
channel, Alice can transfer 0.1 BTC off-chain to Berto: to do so, they
update their balances in the channel accordingly, so Alice's balance is
now 0.4 BTC and Berto's balance is 0.6 BTC. Channels are bidirectional,
so Berto can pay Alice as well. No transactions on the blockchain are
required to accomplish these payments between Alice and Berto: they just
update their bidirectional off-chain balances.

\subsubsection{Payment Network}

A payment network is a connected network of payment channels linked
together. Compared to blockchain transactions, off-chain payments
performed in payment networks are: cheaper, as lower fees are required
to route payments in such networks; faster, as they do not need to be
registered on the blockchain; and more privacy preserving, as they are
not visible in the public blockchain.

In the rest of the section we'll use the following example of a payment
network.

\vspace{6pt}
\noindent {\em Payment network example}. Alice sends 0.1 BTC off-chain
to Davide, without having a direct payment channel with him. She uses a
route of already open channels linking her to Davide: the channel she
has with Berto; the channel between Berto and Carola; and the channel
between Carola and Davide.

\subsection{The Lightning Network}

The Lightning Network (LN) is the implementation of a network of payment
channels for Bitcoin. It was introduced in 2015 by Joseph Poon and
Thaddeus Dryja \cite{ln}.

\subsubsection{LN Payment Channel Lifecycle}

A payment channel lifecycle, as defined by the LN protocol, is presented
in the following.

\vspace{6pt}
\noindent {\em Channel funding}. Alice and Berto create a {\em funding
  transaction}, the transaction required to fund the channel. The
funding transaction is broadcast to the blockchain and, when confirmed,
the channel is opened. Alice and Berto also create a {\em commitment
  transaction} representing their current balances in the channel: such
transaction spends the bitcoins in the funding transaction and transfers
0.5 BTC to Alice and 0.5 BTC to Berto. The commitment transaction is
{\em not} broadcast to \mbox{the blockchain}.

\vspace{6pt}
\noindent {\em Payment execution}. When Alice wants to pay 0.1 BTC to
Berto, the two parties create a new commitment transaction which
reflects the new balances: 0.4 BTC of Alice and 0.6 BTC of Berto. No
transaction is sent to the blockchain, \ie, the operation is executed
off-chain.

\vspace{6pt}
\noindent {\em Channel closing}. When the two parties want to close the
channel, they send the latest commitment transaction to the blockchain.

\vspace{6pt}
\noindent {\em Punishments to disincentivise misbehavior}. Berto can
punish Alice if Alice misbehaves (or vice versa), thanks to the {\em
  Revocable Sequence Maturity Contract} (RSMC), a contract implemented
by a script in the spending condition of the commitment transaction.
Alice could be tempted to broadcast an older commitment to the
blockchain, for instance the commitment preceding her payment to Berto,
when she owned more funds (0.5 instead of 0.4 BTC). If Alice tries to do
so and Berto finds the older commitment transaction on the blockchain,
Berto can withdraw all funds from the channel (including the 0.4 BTC of
Alice's balance), by using the RSMC output contract of the old
commitment transaction broadcast by Alice. In practice, when an older
commitment transaction is replaced by a newer one, Berto is provided
with a special transaction, pre-signed by Alice, that he can
unilaterally broadcast. This special transaction ``invalidates'' the
older commitment transaction as it consumes all the older-commitment
transaction outputs and assigns all channel funds to Berto. Thanks to
this construction, Alice may lose all her funds if caught cheating, and
she is therefore not incentivized to misbehave.

\vspace{6pt}
\noindent {\em Channel unbalancing}. A channel is said unbalanced or
skewed when one balance is much higher than the other. If payments
always flow from Alice to Berto and no payments flow from Berto to
Alice, Alice's balance may go to zero. Channel unbalancing is an issue,
as a payment cannot traverse a channel in a given direction if the
balance in that direction is lower than the \mbox{payment amount}.

\subsubsection{Hashed Timelock Contract (HTLC)}
\label{sec:lnhtlc}

HTLC is a contract for off-chain multi-hop payment through channels that
connect the payment sender and the payment recipient. An off-chain
multi-hop payment is the atomic composition of several off-chain
payments, each one performed on an intermediate payment channel. An
intermediate peer receives an inbound payment on one of its channels and
forwards the same amount to the next payment hop (actually a little less
since it retains an intermediation fee), until the payment reaches the
recipient. The HTLC scheme creates a network of payment channels where
off-chain payments can be routed.

In practice, the HTLC contract is implemented by a script in the
spending condition of the commitment transaction of a payment channel
and it executes conditional payments. Alice pays 0.1 BTC to Berto if
Berto can demonstrate to know a value \R{} within a certain timeout
({\em timelock}, in the Bitcoin jargon): \mbox{in this case}, the HTLC
is fulfilled; otherwise, 0.1 BTC returns back to Alice and the HTLC
fails. Funds in the HTLC payment stay locked up to the disclosure of
\R{} or, if \R{} is never disclosed, up to the timelock's expiration. In
the following, we describe the phases of a payment from Alice to Davide
using HTLCs.

\vspace{6pt}
\noindent {\em HTLC establishment}. For the multi-hop payment originated
by Alice and trying to reach Davide, \mbox{the following} HTLCs are
established: (i) the HTLC between Alice and Berto, where Alice pays 0.1
BTC to Berto if Berto demonstrates to know \R{} within a timelock, \eg,
three days; (ii) the HTLC between Berto and Carola, where Berto pays 0.1
BTC to Carola if Carola demonstrates to know \R{} within a timelock,
\eg, two days; \mbox{ and (3) the HTLC} between Carola and Davide, where
Carola pays 0.1 BTC to Davide if Davide demonstrates to know \R{} within
a timelock, \eg, one day.

\vspace{6pt}
\noindent {\em HTLC fulfillment}. Alice sends \R{} to Davide, and HTLCs
from Davide to Alice are fulfilled: Davide shows \R{} to Carola within
one day, and Carola pays him 0.1 BTC; Carola shows \R{} to Berto within
two days, and Berto pays her 0.1 BTC; Berto shows \R{} to Alice within
three days, and Alice pays him 0.1 BTC. At the end, 0.1 BTC has been
transferred from Alice to Davide. For their work of forwarding the
payment, Berto and Carola can withhold a small fee when transferring the
payment (which we omitted in this example for the sake of simplicity).
It is important to highlight that all the described operations are
performed off-chain, without the need to interact with the blockchain.
Finally, it is worth noticing that, from Davide to Alice, HTLCs have
increasing timelocks. This ensures that each party has enough time to
know \R{} and claim her funds: Davide shows \R{} to Carola in one day,
Carola pays 0.1 BTC to Davide; after that, Carola still has one day to
claim 0.1 BTC from Berto, as the timelock with Berto is set to two days.

\vspace{6pt}
\noindent {\em HTLC failure}. If \R{} is not disclosed by an
uncooperative peer, the payment fails in all channels. \mbox{A failure}
message is sent back to Alice and no payment occurs between the
channels, as HTLC conditions have not been fulfilled. In such case funds
in the HTLC are locked up to the timelock expiration. If Davide does not
disclose \R{}, Carola has to wait one day to get back 0.1 BTC according
to the HTLC conditions. After that, Carola closes the channel with
Davide by broadcasting the last commitment transaction to the blockchain
and propagates back the failure to Alice.

\subsection{Brief Literature}
Although the most developed, the Lightning Network is not the only
off-chain payment network in literature. In 2015 Christian Decker
proposed Duplex Micropayment Channels for building an off-chain payment
network based on HTLCs \cite{decker}. Compared to this approach, the key
innovation in LN is in punishing a misbehaving party. Raiden
\cite{raiden} uses smart contracts to implement the same fundamental
concepts of LN on the Ethereum blockchain. Sprites \cite{sprites} is an
attempt to improve both LN and Raiden, as it aims at minimising the time
during which funds are locked while being transferred via HTLCs.

\subsection{Related Work}
There are currently two publicly available projects \cite{simulatingLn,
  dirouting} labelled as Lightning Network simulators. However, they do
not simulate the HTLC protocol: unlike \htlcsim, they do not present a
complete mapping of Lightning Network code (\cfr{}
Section~\ref{computation-flow}). They are simulators of generic networks
where payments are routed: their purpose is to study specific aspects of
the protocol, and they make no claim of generality nor completeness. To
the best of our knowledge, \htlcsim{} is the first publicly available
HTLC payment network simulator.

\section{Analysis}
\label{sec:analysis}

For implementation level details of HTLC mechanisms, we took as
reference \texttt{lnd}, \mbox{the Golang} implementation of the
Lightning Network. \texttt{lnd}, as the other LN implementations (\eg,
\mbox{\texttt{c-lightning} \cite{c-lightning}}, and \texttt{eclair}
\cite{eclair}), fully conforms to the so-called {\em Basis of Lightning
  Technology} \mbox{(BOLT) \cite{bolt}}, the Lightning Network
specifications.

\begin{figure}[!ht] \centering
\includegraphics[width=1.0\columnwidth]{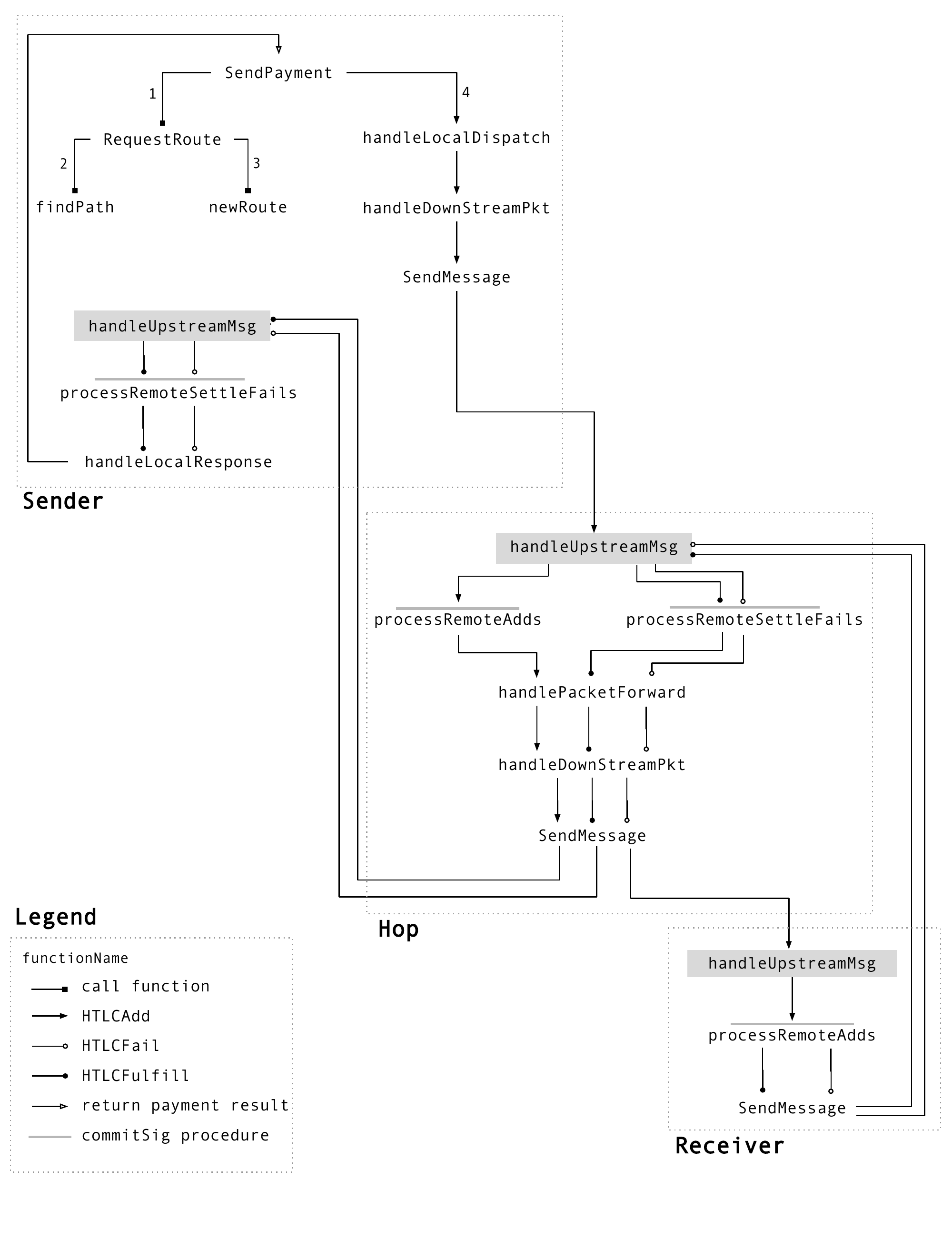}
\caption{Multi-hop payment call graph of \texttt{lnd}.}
\label{fig:callgraph}
\end{figure}

In Figure \ref{fig:callgraph}, we show the multi-hop payment call graph
resulting from our analysis of the release
\href{https://github.com/lightningnetwork/lnd/releases/tag/v0.5-beta}{\texttt{lnd-v0.5-beta}}.
The call graph represents the main functions called when a payment flows
from a sender to a receiver through an intermediate hop. These functions
implement the HTLC mechanism that ensures the atomicity of the payment
(\cfr{} Section \ref{sec:lnhtlc}).

We describe three peers, as they call different functions with different
behaviors: the payment sender \node{A}, the payment receiver \node{D}
and a generic intermediate hop \node{H}. This case can be easily
generalised to n-intermediaries since the behaviour of each intermediate
hop is always the same.

HTLC messages (\ie, \msg{HTLCAdd}, \msg{HTLCFulfill} and \msg{HTLCFail})
that flow through peers are represented by arrows:
\begin{itemize}
\item \msg{HTLCAdd}: sent from \node{A} to \node{D} through \node{H},
  for the establishment of HTLCs among the peers;

\item \msg{HTLCFulfill}: sent from \node{D} to \node{A} through
  \node{H}, for the fulfillment of the HTLCs; and

\item \msg{HTLCFail}: sent in case of failures, for failing the HTLCs.
\end{itemize}

The function \lnd{handleUpStreamMsg} in the gray box represents the
first function called by a peer when an HTLC message is received. The
gray line represents the \lnd{commitSig} procedure that establishes an
HTLC: the creation of a new commitment transaction containing the HTLC
between the two \mbox{involved parties}.

In Appendix \ref{sec:lndcode}, we describe details of the \texttt{lnd}
code and of the functions of the call graph, while in the following we
provide an high-level description of the flow of a payment.

\vspace{6pt}
\noindent {\em Payment initialisation and sending}. Source routing is
run by \node{A} to search for an available route to \node{D}. If~no
route is found, the payment is considered failed; otherwise, \node{A}
sends an \msg{HTLCAdd} message to \node{H} to establish the HTLC.

\vspace{6pt}
\noindent {\em Payment relay (intermediate hop)}. \node{H} checks for
possible errors (\eg, it checks whether it has enough balance to forward
the payment). If there are errors, it sends an \msg{HTLCFail} message
back to \node{A}, failing the HTLC; otherwise, it sends an \msg{HTLCAdd}
message to \node{D}. (This step occurs multiple times in the general
case of a payment going through multiple intermediate hops.)

\vspace{6pt}
\noindent {\em Payment reception}. \node{D} checks for possible errors.
If there are errors, it sends an \msg{HTLCFail} message to \node{H}
(which in turn propagates it to \node{A}), failing the HTLCs; otherwise,
it sends an \msg{HTLCFulfill} message to \node{H} (which in turn
propagates it to \node{A}), for fulfilling the HTLCs.

\vspace{6pt}
\noindent {\em Payment re-attempt}. If an \msg{HTLCFail} is received by
\node{A}, \node{A} tries to re-attempt the payment by searching for a
new route. If the failure was due to a channel which did not have enough
balance to forward the payment, the channel is blacklisted. If the
failure, instead, was due to an uncooperative peer, the peer is
blacklisted. A new route is searched, excluding blacklisted peers and
channels. If no new route is found, the payment is considered
definitively failed.

\section{Design}
\label{sec:design}

In this Section, we provide a detailed description of the \htlcsim{}
simulator design, which is, to the best of our knowledge, the first
simulator for HTLC payment networks. By HTLC payment network, we mean a
network where peers are connected by payment channels and off-chain
payments are routed using HTLC contracts.

The simulator takes as input a definition of HTLC network and payment
script to be played during simulation. It simulates payments in the HTLC
network by locally running a discrete-event mapping of the \texttt{lnd}
code. It produces performance measures in the form of payment-related
statistics (\eg, the probability of payment failures and the mean
payment \mbox{complete time}).

\subsection{Assumptions}
\label{sec:assumpt}

An HTLC payment network relies on a blockchain as a securing mechanism
for all its payment channels. The underlying blockchain is therefore a
fundamental prerequisite of each HTLC payment network. We assume
channels are loaded with an amount of native tokens from the underlying
blockchain and fees for payment forwarding are due to the intermediary
hops in the same denomination.

\htlcsim{} does not consider blockchain interactions during a simulation
execution, since the performance measures produced by the simulator are
related to payments, which are completely performed off-chain.
Simulations run on a network in which no new channels are opened.
\mbox{This condition} is implicitly guaranteed if the whole simulation
time is shorter than the time required to fund a new channel with an
on-chain Bitcoin transaction. An on-chain Bitcoin transaction is usually
considered final after 6 confirmations, statistically 50 minutes after
the first confirmation. This implies a relationship between simulation
duration and likelihood of simulation results: the shorter the
simulation, the more plausible the results. Without loss of generality,
we do not take into account those channels that might have been
established early enough before the simulation starting time to become
operational during the simulation time window. With simulations of
around 15 minutes we obtain a good compromise between meaningfulness of
the experiment and accuracy of results.

Another assumption is that each peer has a complete and precise
knowledge of all other peers and channels in the network. In reality, as
specified in BOLT, new channels and peers are announced through a gossip
protocol. Therefore, real performance may be slightly worse than the one
measured by \htlcsim, as peers may have a slightly imprecise knowledge
of the network due the nature of the gossip protocol.

\subsection{Formal Model}

Having a network with $N$ nodes connected by payment channels and having
$M$ payments to be executed on this network, the simulator formal model
can be expressed using the following formulas:

\begin{equation}
  \label{eq:find}
  \rt = \phi(\pay, \graph_\payindex, \bl)
\end{equation}
\begin{equation}
  \label{eq:move}
  \graph_{\payindex+1} = \mu(\pay, \rt)
\end{equation}

The output $\rt$ is the route found for the payment $\pay$. Function
$\phi$ is the function which finds a route for a payment. It takes as
input the following parameters:
\begin{itemize}
\item $\pay = (n_{i_\payindex}, n_{j_\payindex}, a_\payindex)$ for
  $\payindex = 1, 2, ..., M$ is the payment for which a route has to be
  found: $n_{i_\payindex}$ and $n_{j_\payindex}$ are the sender node and
  the receiver node of the payment, for $i, j = 1, 2, ..., N$ and $i\neq
  j$; and $a_\payindex$ is the payment amount.
\item $\graph_\payindex$ is the graph of nodes $n_i$ for $i = 1, 2, ...,
  N$ connected by payment channels. The subscript $\payindex$ indicates
  the state of the graph at the moment in which $p_k$ is processed, with
  the available nodes and channels and their attributes at that moment.
\item $\bl$ is the blacklist of possible nodes and channels excluded when
  searching for a route for $\pay$ (\mbox{cf. Section \ref{sec:analysis}}).
\end{itemize}

Given payment $\pay$ and the route $\rt$, function $\mu$ executes the
payment $\pay$ along the route $\rt$. $\mu$ function produces a new
state of the graph, namely $\graph_{\payindex+1}$, because the payment
execution causes some changes in the channels involved in the payment,
such as the update of channel balances according to payment amount
$a_\payindex$. A whole simulation is therefore the transition from
$\graph_{0}$ to $\graph_{|M|+1}$, the initial state of the network and
the final one (\ie{}, the state reached after the execution of the last
payment), respectively.

\subsection{Software Architecture}

\htlcsim{} is a {\em discrete-event} simulator \cite{batchmeans}. Events
represent state changes of the simulated system. The {\em event loop} is
the core of the simulation engine: it extracts the next event from a
queue where events are sorted according to their occurring time; it
advances the simulation time to the instant of occurrence of the event
being currently processed; and it calls the function that processes the
event.

\begin{figure}[b]
   \centering
\includegraphics[width=\columnwidth]{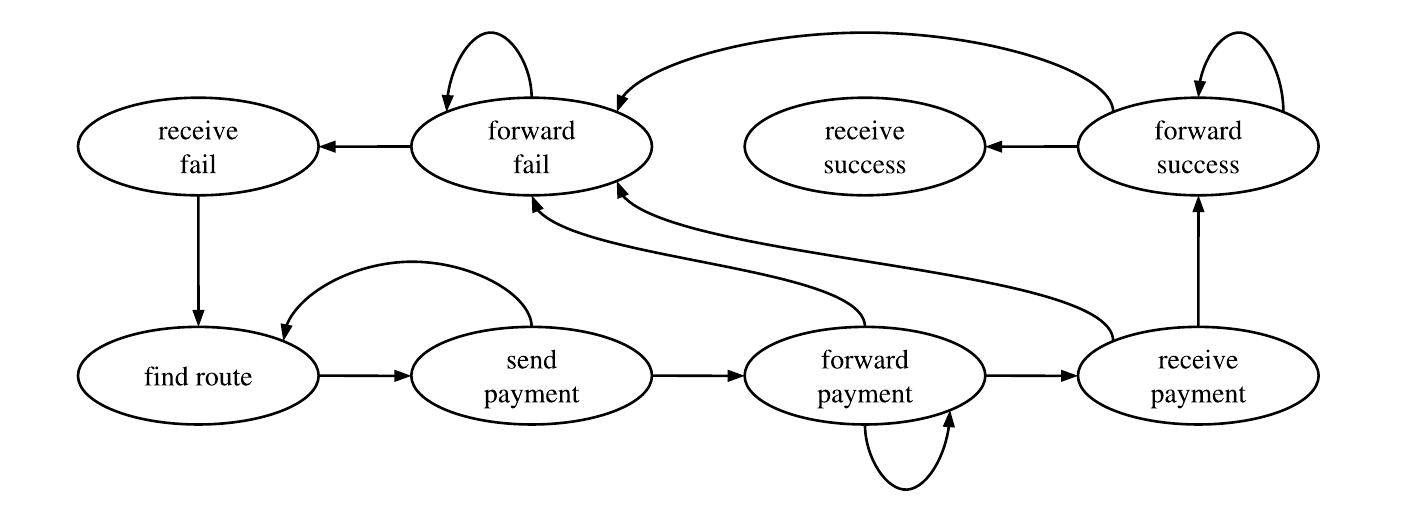}
\caption{\htlcsim{} simulator events state diagram.}
\label{fig:state}
\end{figure}

The simulator generates an event each time a payment changes its state
according to the state diagram in Figure \ref{fig:state}. Each type of
event in the diagram is processed by a function of the same name of the
simulator, as explained in the following Section.

\subsubsection{Computation Flow}
\label{computation-flow}
Table \ref{tab:sim-function} shows the mapping between each function of
the \texttt{lnd} computation flow and functions of the \htlcsim{}
simulator codebase. The completeness of the mapping ensures the validity
of the \mbox{simulated results}.

\begin{table}[!ht]
  \caption{Mapping between \htlcsim{} simulator functions and \texttt{lnd} functions.}
  \label{tab:sim-function}
  \footnotesize
  \begin{tabularx}{\columnwidth}{p{0.18\columnwidth}p{0.18\columnwidth}Xp{0.08\columnwidth}p{0.1\columnwidth}}
  \toprule
  \textbf{Simulator Function} & \textbf{Description} & \textbf{Simulated Functions} & \textbf{Peer} & \textbf{Message} \\
  \midrule  
  \simul{find\_route} & Search for a payment route &   \lnd{SendPayment}, \lnd{RequestRoute}, \lnd{findPath}, \lnd{newRoute} & Sender & - \\
\midrule
  \simul{send\_payment} & Send a payment & \lnd{handleLocalDispatch}, \lnd{handleDownStreamPkt}, \lnd{SendMessage} & Sender & \msg{HTLCAdd} \\
\midrule
  \simul{forward\_payment} & Forward a payment & \lnd{handleUpstreamMsg},
  \lnd{processRemoteAdds}, \lnd{handlePacketForward}, \lnd{handleDownStreamPkt}, \lnd{SendMessage} & Hop & \msg{HTLCAdd} \\
\midrule
  \simul{receive\_payment} & Receive a payment & \lnd{handleUpstreamMsg}, \lnd{processRemoteAdds}, \lnd{SendMessage} & Receiver & \msg{HTLCAdd} \\
\midrule
  \simul{forward\_success} & Forward the successful result of a payment &  \lnd{handleUpstreamMsg}, \lnd{processRemoteSettleFails},
  \lnd{handlePacketForward}, \lnd{handleDownStreamPkt}, \lnd{SendMessage} & Hop & \msg{HTLCFulfill} \\
\midrule
  \simul{forward\_fail} & Forward the fail result of a payment & \lnd{handleUpstreamMsg}, \lnd{processRemoteSettleFails}, \lnd{handlePacketForward}, \lnd{handleDownStreamPkt}, \lnd{SendMessage} & Hop & \msg{HTLCFail} \\
\midrule
  \simul{receive\_success} & Receive the successful result of a payment & \lnd{handleUpstreamMsg}, \lnd{processRemoteSettleFails}, \lnd{handleLocalResponse} & Sender & \msg{HTLCFulfill} \\
\midrule
  \simul{receive\_fail} & Receive the fail result of a payment &  \texttt{handleUpstreamMsg}, \texttt{processRemoteSettleFails}, \texttt{handleLocalResponse} & Sender & \msg{HTLCFail} \\

 \bottomrule
\end{tabularx}
\end{table}

Table \ref{tab:sim-function} makes a clear distinction between specific
behaviors of each \texttt{lnd} function which depends on two parameters:
(i) the type of peer that invokes it (\ie, sender, receiver or
intermediate hop); and (ii) the type of the triggering message
(\msg{HTLCAdd}, \msg{HTLCFail}, \msg{HTLCFulfill}). For example,
\mbox{the function} \simul{send\_payment} of the simulator simulates the
functions \lnd{handleLocalDispatch}, \lnd{handleDownStreamPkt} and
\lnd{SendMessage} of the \texttt{lnd} code, when they are called by the
payment sender in the case of \msg{HTLCAdd} message.

\vspace{6pt}
\noindent {\em Uncooperative behaviour}. Functions
\lnd{forward\_payment}, \lnd{receive\_payment} and
\lnd{forward\_success} simulate the uncooperative behaviour of a peer.
Using Bernoulli probability distributions, we determine whether a peer
is uncooperative and, if so, whether it is uncooperative after or before
establishing the HTLC. If the peer is cooperative, the function is
normally executed. If the peer is uncooperative before establishing the
HTLC, as specified by the LN protocol, an \msg{HTLCFail} is propagated
back to the payment sender and the payment is re-attempted with a new
route that does not involve the uncooperative peer. Finally, if the peer
is uncooperative after establishing the HTLC, as specified by the LN
protocol, an \msg{HTLCFail} is propagated back to the payment sender
after the timelock expiration and the channel connecting to the
uncooperative peer is closed. Then, the sender can re-attempt the
payment, with a new route excluding the closed channel.

\subsubsection{Data Structures}

\begin{figure}[!ht]
   \centering
\includegraphics[width=\columnwidth]{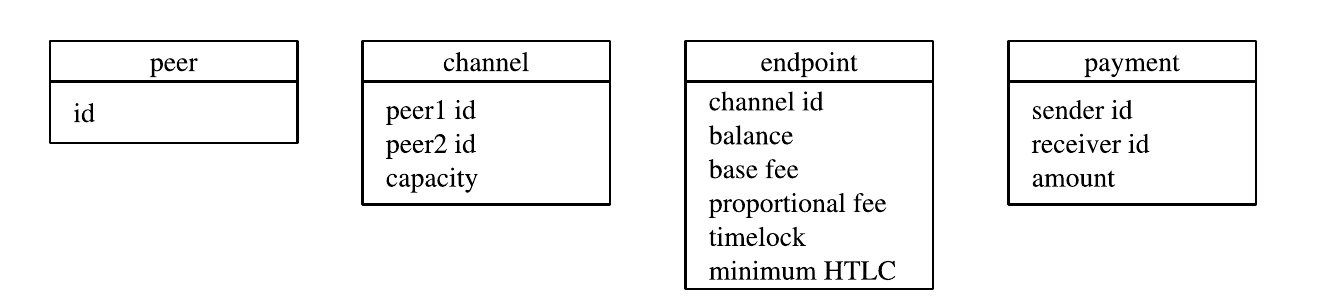}
\caption{\htlcsim{} simulator data structures.}
\label{fig:data-struct}
\end{figure}

Figure \ref{fig:data-struct} shows main attributes of the simulator data
structures. A \simul{channel} connects two \simul{peer} (each one
represented by an ID) and has a certain capacity. The \simul{endpoint}
of a channel describes how a peer behaves in a specific channel and
contains: the current balance of the endpoint in that channel; base and
proportional fee, which constitute the fee withheld by the endpoint for
forwarding a payment (proportional fee is the part of the fee which
depends on the payment amount, while base fee is the constant fee
applied regardless of the payment amount); the timelock set in the HTLCs
established by the channel endpoint; and the minimum value of the HTLC
the endpoint accepts to forward. A \simul{payment} is described by a
sender, a receiver and the payment amount.

\section{Usability}
\label{sec:usability}

This Section describes the usability of \htlcsim; a detailed description
of input/output of a simulation run is provided along with documentation
of the three simulation phases; simulator performances are also
discussed.

\subsection{Workflow}
\label{ssec:workflow}

The workflow to interact with the \htlcsim{} simulator is shown in
Figure \ref{fig:workflow}. It is made of three phases: (i)
pre-processing phase; (ii) simulation phase; and (iii) post-processing
phase. For the sake of clarity, we first introduce the simulation phase.
An explanation of the pre-processing and post-processing phases will
follow.

\subsubsection{Simulation Phase}
\label{sec:sim-stage}

\begin{figure}[!ht]
  \centering
  \includegraphics[width=\columnwidth]{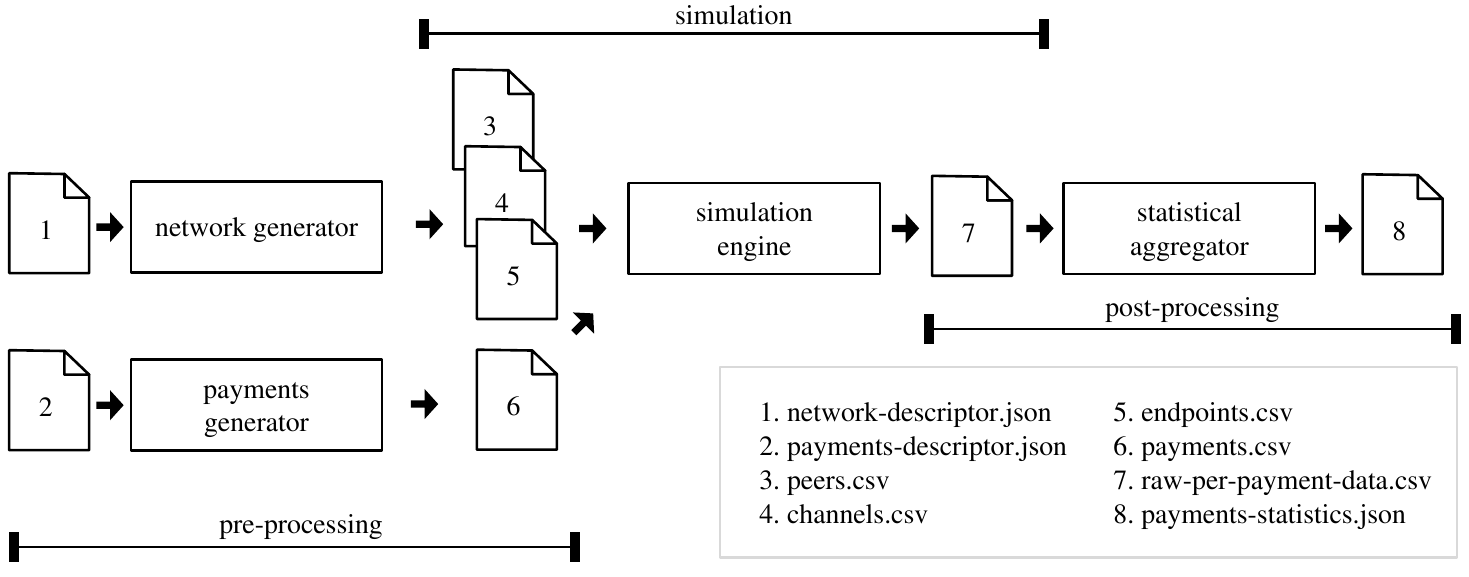}
  \caption{\htlcsim{} simulator workflow.}
  \label{fig:workflow}
\end{figure}

\noindent {\em Input attributes}. The simulation engine requires two
inputs:

\begin{enumerate}
\item A complete specification of the HTLC network (Files 3, 4 and 5 in
  Figure~\ref{fig:workflow}). These files allow for a fine grained
  specification of the attributes of Figure \ref{fig:data-struct} of
  peers (\texttt{peers.csv}), channels (\texttt{channels.csv}) and
  endpoints (\texttt{endpoints.csv}).
\item The detailed list of payments to be executed with their attributes
  (File 6 in Figure~\ref{fig:workflow}, \texttt{payments.csv}),
  specifying sender, receiver, payment amount and triggering time.
\end{enumerate}

\noindent Although verbose, this input specification allows for the most
detailed definition of the simulation scenario. For a succinct and
statistically meaningful input configuration mode, the reader can resort
to the pre-processing phase of the workflow described below.

\vspace{8pt}
\noindent {\em Simulation engine execution}. Given the HTLC network and
the payments specified as input, \mbox{the simulation} engine executes
the payments on the HTLC network, running the actual discrete-event
simulation.

\vspace{6pt}
\noindent {\em Output attributes}. The simulation phase outputs detailed
information about each simulated payment (File 7 in
Figure~\ref{fig:workflow}, \texttt{raw-per-payment-data.csv}):
\begin{itemize}
\item payment end time;
\item payment result: success or failure;
\item number of times payment was attempted;
\item whether payment encountered an uncooperative peer; and
\item route traversed by the payment, if present.
\end{itemize}

\subsubsection{Pre-Processing Phase}

We wanted the \htlcsim{} simulator to allow fine-grained simulations as
well as the option to concisely describe a simulation from a statistical
point of view. While the input described in Section~\ref{sec:sim-stage}
directly supports the first case, the second requirement is fulfilled by
the pre-processing phase. This phase takes as input a few parameters
that statistically define both the network and the payments to be
simulated. Then, it randomly generates an instance of HTLC network and
an instance of payment script that match the input description.

\vspace{6pt}
\noindent {\em Input parameters}. Input parameters of the pre-processing
phase with their symbols are shown in \mbox{Table \ref{tab:input}}. The
following parameters deserve an explanation:
\begin{itemize}
\item \sigmatop{} tunes the presence of hubs in the network topology. It
  is the width of the Gaussian distribution defining the probability of
  connection among peers. For each peer, this Gaussian probability
  distribution is used for choosing the other endpoint of one of the
  peer channels. Therefore, if the width of this Gaussian is zero, all
  peers have one channel open with the same peer, which, consequently,
  will be a hub. On the other hand, if the Gaussian width is infinite
  any peer has the same probability to be connected to any other, thus
  producing a totally \mbox{decentralised network}.
\item \sigmaam{} tunes payment amounts. It is the width of the Gaussian
  distribution whose tail is used to choose the orders of magnitude of
  payment amounts. The greater this width is, the higher the payment
  amounts.
\item \samedest{} is the fraction of the total payments directed toward
  the same recipient. This parameter allows to model the use case of
  many small payments sent to the same destination peer, \eg{}, to a
  provider of video-streaming services which is paid for each short
  segment of video streaming.
\end{itemize}

\begin{table}[H]
\centering
\caption{\htlcsim{} simulator pre-processing phase input parameters.} \label{tab:input}
\begin{tabular}{ccl}
\toprule
\textbf{Type} & \textbf{Symbol} & \textbf{Definition}  \\
\midrule
\multirow{7}{*}{Network} & \npeers & Number of peers  \\
& \nchannels & Average number of channels per peer \\
& \sigmatop & Tuner of network topology  \\
& \uncoopbefore & Uncooperative peers probability before HTLC establishment  \\
& \uncoopafter & Uncooperative peers probability after HTLC establishment  \\
& \capacity & Average channel capacity  \\
& \gini & Gini index of channel capacity \\
\midrule
\multirow{4}{*}{Payments} & \payrate & Transactions per second (off-chain) \\
& \npay & Number of payments  \\
& \sigmaam &  Tuner of payment amounts \\
& \samedest & Fraction of same-recipient payments \\
\bottomrule
\end{tabular}
\end{table}

\subsubsection{Post-Processing Phase}

The post-processing phase transforms raw per-payment simulation output
attributes into statistically meaningful performance measures (File 8 in
Figure \ref{fig:workflow}, \texttt{payments-statistics.json}). To do so,
it applies the {\em batch means} method \cite{batchmeans}, which allows:
(i) producing performance results that are not influenced by the initial
transient state, where the system is not stable; and (ii) computing
statistical mean, variance and 95\% confidence interval for each
measure. The {\em batch means} method consists in dividing a simulation
run into multiple batches, which are statistically independent among
each other. Output measures are zeroed and re-computed at each batch.
Each final output measure is the statistical mean of that measure over
the batches and comes also with variance and 95\% confidence interval.

\vspace{6pt}
\noindent {\em Simulation results format}. The performance measures
produced by the simulator are shown in Table \ref{tab:output} with their
symbols. Some clarifications are presented in the following:
\begin{itemize}
\item \proute{} is the probability that a payment fails due to the
  absence of a route connecting sender and~receiver.
\item \pbalance{} is the probability that a payment fails because a
  channel in the route was unbalanced and an alternative route is not
  found.
\item \puncoop{} is the probability that a payment fails because a peer
  in the route was uncooperative and an alternative route is not found.
\item \ptimeout{} is the probability that a payment fails because it
  took more than a timeout of 60 seconds to be completed.
\item \punk{} represents the remaining fraction of payments for which we
  do not know whether they failed or succeeded, as they ended after the
  time validity window of our simulation. For example, this category can
  encompass payments delayed after a long timelock, as a peer was
  uncooperative after establishing the HTLC for the payment.
\item \paytime{} is the mean time for a successful payment to complete.
\item \attempts{} is the mean number of times a payment is re-attempted.
\item \route{} is the mean route length traversed by a successful
  payment.
\end{itemize}

\begin{table}[H]
\centering
\caption{\htlcsim{} simulator performance measures.} \label{tab:output}
\begin{tabular}{cl}
\toprule
\textbf{Symbol} & \textbf{Definition} \\
\midrule
\psucc & Probability of payment success   \\
\proute & Probability of payment failure for no route  \\
\pbalance & Probability of payment failure for unbalancing   \\
\puncoop & Probability of payment failure for uncooperative peers  \\
\ptimeout & Probability of payment failure for timeout expiration \\
\punk & Probability of unknown payments \\
\paytime & Payment complete time  \\
\attempts & Number of payment attempts  \\ 
\route & Payment route length  \\
\bottomrule
\end{tabular}
\end{table}

\subsection{Performance Analisys}
\label{ssec:performance}

Table~\ref{tab4} shows the performance of the \htlcsim{} simulator in
terms of execution time. Simulation times refer to experiments run on a
machine of model DGX-1, manufactured by Nvidia, located in Rome, Italy.
The machine is equipped with 80 CPUs of model Intel\textsuperscript{®}
Xeon\textsuperscript{®} CPU E5-2698 v4 @ 2.20 GHz and 512 GB of RAM.
Table~\ref{tab4}a shows that execution time increases with the increase
of the number of peers of the simulated network. This is due to the fact
that time spent by the Dijkstra's algorithm, used to find payment
routes, grows with the number of peers. With 1 million peers, a
simulation run requires more than four days.

\begin{table}[H]
      \centering
        \begin{tabular}{cc}
            \toprule
           \multicolumn{2}{c}{ (\textbf{a})}\\
            \midrule
            \textbf{Peers} & \textbf{Execution Time (h)} \\
            \midrule
            100,000 & 6.64 \\
            200,000 & 15.92 \\
            500,000 &  49.25 \\
            700,000 & 64.68 \\
            1,000,000 & 104.15 \\
        \midrule
            \multicolumn{2}{c}{(\textbf{b})}\\
            \midrule
          \textbf{Dijkstra calls} & \textbf{Execution Time (h)}\\
          \midrule
          58927 & 4.98 \\
          303374 & 28.3 \\
          \bottomrule
      \end{tabular}
    \caption{\htlcsim{} simulator execution time.}
  \label{tab4}
\end{table}

Table~\ref{tab4}b shows that the execution time also increases with the
increase of the number of calls to the Dijkstra's algorithm. An
execution of the Dijkstra's algorithm is required each time a payment is
attempted. As a payment is re-attempted, calls to the Dijkstra's
algorithm grow as well al simulation execution time. Execution of the
Dijkstra's algorithm is a simulator performance bottleneck. However,
there are no limits to the values of input parameters. The only effect
of setting high values (\eg, an high number of peers or payments) is a
correspondingly long execution time.

\section{Conclusions}
\label{sec:conclusions}
The present research work focused on the Lightning Network, the
mainstream proposal that aims to address the well-known scalability
problem of the Bitcoin blockchain. This solution implements an HTLC
payment network, \ie, a network to securely route off-chain payments
through \mbox{HTLC contracts}.

We developed \htlcsim, a simulator of HTLC payment networks. The
simulator takes as input: \mbox{(i) parameters} representing the HTLC
network to be studied (\eg, peers, channels, \mbox{ and channel}
capacities); and (ii) parameters defining characteristics of the
payments to be simulated on the HTLC network (\eg, payment rate and
payment amounts). \htlcsim{} simulates the defined payments script on
the input network and produces performance measures in the form of
payment-related statistics (\eg,~the probability of payment failures and
the mean time to complete a payment).

Since LN is an emergent network with no central coordination, it's
important to understand in advance what's needed to guarantee healthy
network growth, that is quick and successful payments. We consider
\htlcsim{} a useful predicting tool to simulate possible troublesome
configurations and to check in advance the effects of a routing
optimisation action (\eg, a node that joins the network, and then
establishes and funds several payment channels).

\subsection{Future work}
\label{ssec:future-work}
We plan to follow \texttt{lnd} code and LN mainnet changes with regular
\htlcsim{} upgrades, and LN mainnet snapshots. We'll then simulate
various payment scripts on such snapshots, in order to understand how LN
mainnet might respond to different payment dynamics. We'll also simulate
different networks, to explore how the network could and should evolve.

To better understand the impact of new LN features, in future works we
plan to simulate features that have been proposed but are not yet part
of LN specifications (\eg{}, Atomic Multi-Path Payments \cite{amp},
Split Payments \cite{piatkivskyi2018split}). Conversely, a bottom-up,
data-driven approach could lead to empirically find desirable LN
features which have not been proposed yet.

Finally, we plan a complete exploration of the whole space of meaningful
LN configurations. To do so we'll need to improve \htlcsim{} performance
and run simulations with millions of nodes in reasonable times. A
performance bottleneck we will address is the inherently sequential
application of the Dijkstra's algorithm.


\vspace{6pt} 

\acknowledgments{We thank Prof. Michele Garetto, who gave us a
  fundamental help in understanding the theoretical concepts of
  simulations.}

\abbreviations{The following abbreviations are used in this manuscript:\\

\noindent 
\begin{tabular}{@{}ll}
  LN & Lightning Network \\
  HTLC & Hashed Timelock Contract
\end{tabular}}

\appendixtitles{yes}

\appendixsections{multiple}

\appendix

\section{Reference Code Explanation}
\label{sec:lndcode}

In this section, we describe the details of the \texttt{lnd} code we
took as reference to develop \htlcsim. In \texttt{lnd}, there are two
main code structures which manage payments: \lnd{Switch} and \lnd{Link}.
\lnd{Switch} is the messaging bus of HTLC messages: it is in charge of
forwarding HTLCs or redirecting HTLCs initiated by the local peer to the
proper functions. \lnd{Link} is the service which drives a channel
commitment update procedure according to the HTLCs that concern the
channel. In the following, we describe the functions included in the
call graph in Figure \ref{fig:callgraph}.

\begin{itemize}
\item \lnd{SendPayment}. As shown by the numbers in the arrows, which
  represent the order of function calls, this function first tries to
  find possible routes to transfer the payment to the receiver and then
  tries to send the payment through one of the routes found. If the
  payment fails, it re-attempts the payment through another viable
  route.

\item \lnd{RequestRoute}. It attempts to find candidate routes which can
  route the payment to the receiver.

\item \lnd{findPath}. It runs the Dijkstra's algorithm, using timelock
  and fee as distance metric. In fact, each channel endpoint has a {\em
    policy} which defines the timelock and the fee that will be applied
  to any HTLC forwarded by that endpoint. The higher are the timelock
  and the fees in the endpoint policy, \mbox{the higher} \mbox{the
    distance}.

\item \lnd{newRoute}. It attempts to transform a path into a {\em
    route}. A route is a path which connects sender and receiver and
  which can also transfer the payment. A path is considered capable of
  transferring a payment if all channels in the path have a capacity
  greater than or equal to the payment amount, taking fees into account.
  A fee is the amount of funds a channel endpoint withholds for
  forwarding a payment through that channel.

\item \lnd{handleLocalDispatch}. Function of the \lnd{Switch} that
  processes HTLCs relative to payments initiated by the local peer. This
  function returns an error if there are no channels to the next hop
  with enough balance to forward the payment.

\item \lnd{handleLocalResponse}. Function of the \lnd{Switch} that
  processes the result of a payment initiated by the local peer. It
  receives an \msg{HTLCFail} or \msg{HTLCFulfill} message and it
  propagates them back to \lnd{SendPayment}.

\item \lnd{handlePacketForward}. Function of the \lnd{Switch} that
  processes HTLCs of payments initiated by other peers and to be
  forwarded by the local peer. It produces an \msg{HTLCFail} if no
  channel to the next route hop has enough balance to forward the
  payment, or if the local peer policy is not respected. The
  \msg{HTLCFail} message is then sent back to the payment sender.

\item \lnd{handleDownStreamPkt}. Function of the \lnd{Link} which
  processes HTLCs coming from the \lnd{Switch}. \mbox{It produces} an
  \lnd{HTLCFail} if the channel does not have enough balance to forward
  the payment.

\item \lnd{handleUpStremMsg}. Function of the \lnd{Link}, the first
  called by a peer upon the reception of an HTLC message by another
  peer.

\item \lnd{SendMessage}. Function to send over the network an HTLC
  message from a peer to another.

\item \lnd{processRemoteAdds}. It processes an \lnd{HTLCAdd} and checks
  for possible errors. When called by the payment receiver, the function
  decides whether to accept or not the payment.

\item \lnd{processRemoteSettleFails}. It processes an \lnd{HTLCFail} or
  \lnd{HTLCFulfill} and checks for possible errors.
\end{itemize}

\reftitle{References}
\externalbibliography{yes}
\bibliography{bibliography}

\begin{thebibliography}{}

\bibitem[\protect\citeauthoryear{??}{c-l}{}]{c-lightning}
c-lightning.
\newblock Available online: \url{https://github.com/ElementsProject/lightning}
  (accessed on 31 July 2018).

\bibitem[\protect\citeauthoryear{??}{ecl}{}]{eclair}
eclair.
\newblock Available online: \url{https://github.com/ACINQ/eclair} (accessed on
  31 July 2018).

\bibitem[\protect\citeauthoryear{??}{bol}{}]{bolt}
Lightning network specifications.
\newblock Available online:
  \url{https://github.com/lightningnetwork/lightning-rfc} (accessed on 31 July
  2018).

\bibitem[\protect\citeauthoryear{??}{pay}{}]{paychannel}
Payment channels.
\newblock Available online: \url{https://en.bitcoin.it/wiki/Payment_channels}
  (accessed on 4 August 2018).

\bibitem[\protect\citeauthoryear{??}{rai}{}]{raiden}
Raiden network.
\newblock Available online: \url{https://raiden.network/} (accessed on 31 July
  2018).

\bibitem[\protect\citeauthoryear{Bonneau, Miller, Clark, Narayanan, Kroll, and
  Felten}{Bonneau et~al.}{2015}]{sok}
Bonneau, Joseph, Andrew Miller, Jeremy Clark, Arvind Narayanan, Joshua~A Kroll,
  and Edward~W Felten. 2015.
\newblock Sok: Research perspectives and challenges for bitcoin and
  cryptocurrencies.
\newblock In {\em Security and Privacy (SP), 2015 IEEE Symposium on}, pp.\
  104--121. IEEE.

\bibitem[\protect\citeauthoryear{Burchert, Decker, and Wattenhofer}{Burchert
  et~al.}{2017}]{funding}
Burchert, Conrad, Christian Decker, and Roger Wattenhofer. 2017.
\newblock Scalable funding of bitcoin micropayment channel networks.

\bibitem[\protect\citeauthoryear{Conoscenti, Vetrò, De~Martin, and
  Spini}{Conoscenti et~al.}{2018}]{info9090223}
Conoscenti, Marco, Antonio Vetrò, Juan~Carlos De~Martin, and Federico Spini.
  2018.
\newblock The cloth simulator for htlc payment networks with introductory
  lightning network performance results.
\newblock {\em Information\/}~{\em 9\/}(9).
\newblock
  doi:{\changeurlcolor{black}\href{https://doi.org/10.3390/info9090223}{\detokenize{10.3390/info9090223}}}.

\bibitem[\protect\citeauthoryear{Decker and Wattenhofer}{Decker and
  Wattenhofer}{2015}]{decker}
Decker, Christian and Roger Wattenhofer. 2015.
\newblock A fast and scalable payment network with bitcoin duplex micropayment
  channels.
\newblock In {\em Symposium on Self-Stabilizing Systems}, pp.\  3--18.
  Springer.

\bibitem[\protect\citeauthoryear{Di~Stasi, Avallone, Canonico, and
  Ventre}{Di~Stasi et~al.}{}]{dirouting}
Di~Stasi, Giovanni, Stefano Avallone, Roberto Canonico, and Giorgio Ventre.
\newblock Routing payments on the lightning network.

\bibitem[\protect\citeauthoryear{Gervais, Karame, W{\"u}st, Glykantzis,
  Ritzdorf, and Capkun}{Gervais et~al.}{2016}]{gervaisperformance}
Gervais, Arthur, Ghassan~O Karame, Karl W{\"u}st, Vasileios Glykantzis, Hubert
  Ritzdorf, and Srdjan Capkun. 2016.
\newblock On the security and performance of proof of work blockchains.
\newblock In {\em Proceedings of the 2016 ACM SIGSAC Conference on Computer and
  Communications Security}, pp.\  3--16. ACM.

\bibitem[\protect\citeauthoryear{Jain}{Jain}{1990}]{batchmeans}
Jain, Raj. 1990.
\newblock {\em The art of computer systems performance analysis: techniques for
  experimental design, measurement, simulation, and modeling}.
\newblock John Wiley \& Sons.

\bibitem[\protect\citeauthoryear{Khalil and Gervais}{Khalil and
  Gervais}{2017}]{khalil2017revive}
Khalil, Rami and Arthur Gervais. 2017.
\newblock Revive: Rebalancing off-blockchain payment networks.
\newblock In {\em Proceedings of the 2017 ACM SIGSAC Conference on Computer and
  Communications Security}, pp.\  439--453. ACM.

\bibitem[\protect\citeauthoryear{Miller, Bentov, Kumaresan, and McCorry}{Miller
  et~al.}{2017}]{sprites}
Miller, Andrew, Iddo Bentov, Ranjit Kumaresan, and Patrick McCorry. 2017.
\newblock Sprites: Payment channels that go faster than lightning.

\bibitem[\protect\citeauthoryear{Nakamoto}{Nakamoto}{2008}]{nakamoto}
Nakamoto, Satoshi. 2008.
\newblock Bitcoin: A peer-to-peer electronic cash system.

\bibitem[\protect\citeauthoryear{Osuntokun}{Osuntokun}{}]{amp}
Osuntokun, Olaoluwa.
\newblock Amp: Atomic multi-path payments over lightning.
\newblock Available online:
  \url{https://lists.linuxfoundation.org/pipermail/lightning-dev/2018-February/000993.html}
  (accessed on 31 July 2018).

\bibitem[\protect\citeauthoryear{Piatkivskyi and Nowostawski}{Piatkivskyi and
  Nowostawski}{2018}]{piatkivskyi2018split}
Piatkivskyi, Dmytro and Mariusz Nowostawski. 2018.
\newblock Split payments in payment networks.
\newblock In {\em Data Privacy Management, Cryptocurrencies and Blockchain
  Technology}, pp.\  67--75. Springer.

\bibitem[\protect\citeauthoryear{Poon and Dryja}{Poon and Dryja}{2016}]{ln}
Poon, Joseph and Thaddeus Dryja. 2016.
\newblock The bitcoin lightning network: Scalable off-chain instant payments.

\bibitem[\protect\citeauthoryear{Prihodko, Zhigulin, Sahno, Ostrovskiy, and
  Osuntokun}{Prihodko et~al.}{2016}]{flare}
Prihodko, Pavel, Slava Zhigulin, Mykola Sahno, Aleksei Ostrovskiy, and Olaoluwa
  Osuntokun. 2016.
\newblock Flare: An approach to routing in lightning network.

\bibitem[\protect\citeauthoryear{Reynolds}{Reynolds}{2017}]{simulatingLn}
Reynolds, Diane. 2017.
\newblock Simulating a decentralized lightning network with 10 million users.
\newblock Available online:
  \url{https://hackernoon.com/simulating-a-decentralized-lightning-network-with-10-million-users-9a8b5930fa7a}
  (accessed on 14 December 2018.

\bibitem[\protect\citeauthoryear{Sompolinsky and Zohar}{Sompolinsky and
  Zohar}{2013}]{accelerating}
Sompolinsky, Yonatan and Aviv Zohar. 2013.
\newblock Accelerating bitcoin’s transaction processing.

\bibitem[\protect\citeauthoryear{Vu}{Vu}{}]{lndev}
Vu, Bryan.
\newblock Exploring lightning network routing.
\newblock Available online:
  \url{https://blog.lightning.engineering/posts/2018/05/30/routing.html}
  (accessed on 31 July 2018).

\end{thebibliography}

\end{document}